\RequirePackage{fix-cm}
\documentclass[twocolumn,final]{svjour3}          
\smartqed  
\usepackage{graphicx}
\usepackage{subfigure}
\journalname{Computer Graphics International (CGI) 2012} 
\begin{document}

\title{The Affect of Lifestyle Factors on Eco-Visualization Design}
\subtitle{}
\author{Stephen Makonin \and Maryam H. Kashani \and Lyn Bartram}
\institute{
   S. Makonin \at School of Computing Science \\ Simon Fraser University, Burnaby BC Canada \\ Email: smakonin@sfu.ca 
   \and M.H. Kashani  \at School of Interactive Arts + Technology \\ Simon Fraser University, Surrey BC Canada  \\ Email: mhaghigh@sfu.ca
   \and L. Bartram \at School of Interactive Arts + Technology \\ Simon Fraser University, Surrey BC Canada \\ Email: lyn@sfu.ca
}
\date{ }

\maketitle

\begin{abstract}
As people become more concerned with the need to conserve their power consumption we need to find ways to inform them of how electricity is being consumed within the home. There are a number of devices that have been designed using different forms, sizes, and technologies. We are interested in large ambient displays that can be read at a glance and from a distance as informative art. However, from these objectives come a number of questions that need to be explored and answered. To what degree might lifestyle factors influence the design of eco-visualizations? To answer this we need to ask how people with varying lifestyle factors perceive the utility of such devices and their placement within a home. We explore these questions by creating four ambient display prototypes. We take our prototypes and subject them to a user study to gain insight as to the questions posed above. This paper discusses our prototypes in detail and the results and findings of our user study.
\keywords{eco-visualization \and informative art \and ambient display \and power consumption \and energy conservation \and sustainability}
\end{abstract}

\section{Introduction}
As people become more concerned with the need to conserve their power consumption we need to find ways to inform them of how electricity is being consumed within the home. There is a gambit of research in eco-visualizations (and eco-feedback devices) \cite{6065016,darby2006effectiveness,froehlich2010design,Horn:2011:LSF:1999030.1999051,makonin2011elements,Pierce:2008:EAD:1517744.1517746,rodgersposter,Strengers:2011:DES:1978942.1979252}  all looking at conserving energy consumption in a home. Some focus on hardware devices, some on software/visualizations, and others on the process and effectiveness of designing these devices. 

We are interested in large ambient displays that can be read at a glance and from a distance as informative art (both abstract and pictorial). We want to convey energy consumption information in a way that does not require an understanding of how energy is measured, what those measurement units are, and what those units mean; but, at the same time communicate how much energy an appliance is using compared to other appliances and/or the total home in a relative manner. For instance, if an appliance consumes 50\% of the total energy consumption of a given house, the visualization should show this percentage. The understanding of this percentage should then transfer to the understanding that this also means 50\% of the amount on their energy bill. Our objective is not to teach the consumer what a kWh is (as in other research studies); nor is it to research comprehension.

While understanding comprehension is an important issue, we want to focus on the smaller issues of appropriate usage that affect design and placement. Our main goal is to understand eco-visualization design and localization. Our research question is put as this: \textit{to what degree might lifestyle factors influence the design of eco-visualizations?} To answer this we need to ask how people with varying lifestyle factors perceive the utility of such devices and their placement within a home. 

Our opinion is that \textit{lifestyle factors (age, gender and busyness) all have an influence on the choice of color and form of the ambient display and eco-visualization}. We further conjecture that: gender may have a role in the choice of palette selection; a person's level of busyness may constrain the complexity of design; and persons would want ambient displays in convenient locations.

Our research question and conjectures were explored by creating an ambient display with 4 eco-visualizations that were subjected to an informal user study. Design considerations need to be explored to frame what life\-style and visualization factors are influential before we can administer rigorous scientific method \cite{tory2005doeswork,Greenberg:2008:UEC:1357054.1357074,acevedo2006subjective}.

Reflecting on this, but in more detail, the remainder of our paper is organized as follows. We first look at the related work (Section 2) of what other research has been done in these topic areas. Next, we discuss the visualization design parameters (Section 3) and the design of our prototype system (Section 4) for rendering and visualizing consumption data on an ambient screen.  We change the focus of our paper to the user study we performed (Section 5) and summarize the results (Section 6). We end with a general discussion of our observations and the new hypotheses we have discovered (Section 7) and conclude with future work (Section 8).

\section{Background}
Our research touched on a number of different research areas: ambient displays, eco-visualization, and lifestyle factors. There are a number of factors that we need to consider from each research area that affects our designs. Ambient displays and eco-visualizations come from the area of information visualization and human computer interaction (HCI). Lifestyle factors come from research in architecture (buildings, habitation, and\break spaces). We are attempting to account for this architectural issue in the design and evaluation of our prototype.

\subsection{Ambient Display}
In previous research by Pousman and Stasko, ambient displays have been defined, and a taxonomy is presented along four design dimensions: \textit{Information Capacity}, \textit{Notification Level}, \textit{Representational Fidelity}, and \textit{Aesthetic emphasis} \cite{pousman2006taxonomy}. The two taxonomies relevant to our study are Information Capacity, and Representational Fidelity.

Pousman and Stasko define Information Capacity as the representation of the number of ``discrete information sources that a system can represent'' \cite{pousman2006taxonomy}. They further explain some systems can display a single piece of data such as a Table Fountain (or water lamp). Other systems can display the value of 20 (or more) different information elements on one screen (e.g. Apple Dashboard). Systems are ranked from \textit{Low} to \textit{High}. 

Pousman and Stasko define the representational fidelity as the display components of a system and ``how the data from the world is encoded into patterns, pictures, words, or sounds'' \cite{pousman2006taxonomy}. They explain some systems can be very abstract in their representation of data while others can be very direct. As with the previous taxonomy, systems are ranked from \textit{Low} to \textit{High} with low as something similar to an Ambient Orb, and high as an InfoCanvas.

\subsection{Eco-Visualization}
When we discuss eco-visualizations we also take into account the discussion of eco-feedback devices. We see little difference between eco-visualization and eco-feedback devices only to say that an eco-visualization is one type of eco-feedback device.

Eco-visualization is a relatively new area of research. The main focus for researchers like Makonin, Rodgers, Bartram, and others in this area is the recording of consumption in a house and then visualizing that consumption in a way that is meaningful to a homeowner. What remains the big question in this area is, \textit{what is meaningful, and what does meaningful mean?}. Makonin \cite{makonin2011elements} has previously taken power, water, and natural gas consumption and visualized it as abstract art using fluid dynamics. The visualization that he developed has had positive feedback (informally). However, there was no formal user study performed to verify any part of the prototypes he developed in an attempt to answer the question on \textit{meaningful}. The benefit of his work has been that the data he has used in this paper and in his other work \cite{makonin2011intelligent} is from a real home.

Rogers and Bartram have done extensive work on \textit{human home interaction} and investigating and designing a framework for such devices \cite{rodgersposter}. Bartram has continued to look at issues affecting the ability of homeowners to conserve energy \cite{bartram2011smart1,bartram2011smart2}. Their research reiterates the difficulties for systems to communicate to the homeowner how their house is performing or even what to communicate. This leads to two questions. \textit{What is the right amount of information a homeowner needs? How can we insure that the homeowner (regardless of education) understands the meaning of this information?} We attempt to avoid this by using simplified relative information based on percentage of total household consumption.

Froehlich \cite{froehlich2010design}, Horn \cite{Horn:2011:LSF:1999030.1999051}, Pierce \cite{Pierce:2008:EAD:1517744.1517746}, Strengers \cite{Strengers:2011:DES:1978942.1979252}, and Lilley \cite{6065016} all explore the current and future designs of eco-feedback devices and eco-visualizations. There seems to be a need to study and restudy the devices out there. We believe that the only way to make forward progress is to create new prototype devices and perform user studies on them. By engaging users with different prototypes we think we can gain different insights into the difficult subject of \textit{effective feedback}. We make design decisions based on introspection (designer's intuition), and the observations made in the above research. From the execution of our user study we can then observe whether these designs stand or fall. We can then create new prototypes based on our user study results and execute new user studies (iteratively, of course).

\subsection{Lifestyle Factors}
Household characteristics influence occupant behavior and, therefore, differences in such attributes may lead to significant differences in energy use. Such factors are what we define as lifestyle factors. Lifestyle factors can consist of external environmental factors, household characteristics, and individual detriments and choices \cite{bin2005consumer}.  The difference mainly lies between whether they are \textit{malleable} or \textit{intractable} lifestyle factors \cite{bartram2011smart1}. The former are those that can be influenced, such as psychological characteristics, habit, knowledge, contextual awareness, and convenience. On the other hand, fixed or intractable factors ``are not prone to individual interventions (such as age, gender or income)'' \cite{kashani2011lifestyle}. 

Considering eco-visualizations are mainly used to inform people about their energy consumption, and idealistically motivate them to conserve through better understanding of their energy usage, it is important to design ambient displays with respect to people's characteristics and needs. We are very interested in exploring these \textit{lifestyle factors} of households as they pertain to the design of eco-visualizations. We feel that intractable factors of age, gender, and busyness (which is partly malleable and partly intractable) are factors that have a direct implication as to how these devices are designed. 

\section{Visualization Design}
Our goal was to design visualizations based on a set of explanatory measurements and choices in visualization encoding methods.
The explanatory measurements we choose have direct implications on our choice of visualization encoding methods. This section is devoted to the understanding of this.

\subsection{Explanatory Measurements}
Our research explores the ways in which ambient displays and eco-visualizations reflect the lifestyle of households. We selected three major intractable lifestyle factors as our independent variables that affect the design of the display: \textit{age} and \textit{gender}, with an observation variable called \textit{busyness}. 

Age is considered an intractable lifestyle factor (it is not within the power of the person to change) that would categorize persons in different age groups. Two age categories were decided on: adult (19 to 49) and senior (50 and over). The age category may have an effect on how eco-visualizations are designed.

Gender is generally considered an intractable lifestyle factor that would categorize persons as either male or female. We, again, hypothesize that the gender category has an effect on how eco-visualizations are designed.

Busyness (or level of busyness) is a lifestyle factor that can be argued as either malleable or intractable. Whether it is either one or the other has no bearing on the research in this paper, so we will consider it as both. Busyness is broken down into three categories: busy, fairly busy, and not busy. This is a subjective measure based on how busy a person perceives themselves to be; not whether they are actually busy. We are interested in their perception of busyness because we think it will also affect designs.

We are aware that lifestyle factors also have an effect on how energy is consumed within the household. However, we believe this is an issue that is out of scope and does not affect the design of our eco-visualizations. As we have  discussed, our designs focus on how relative consumption (percent of the total house consumption) is visualized and what the appliances are consuming. The only thing to note is that the lifestyle factors of different households might lead to different appliances being monitored.

\subsection{Visualization Encoding Methods}
\begin{figure}[ht]
\centering
\includegraphics[width=0.9\columnwidth]{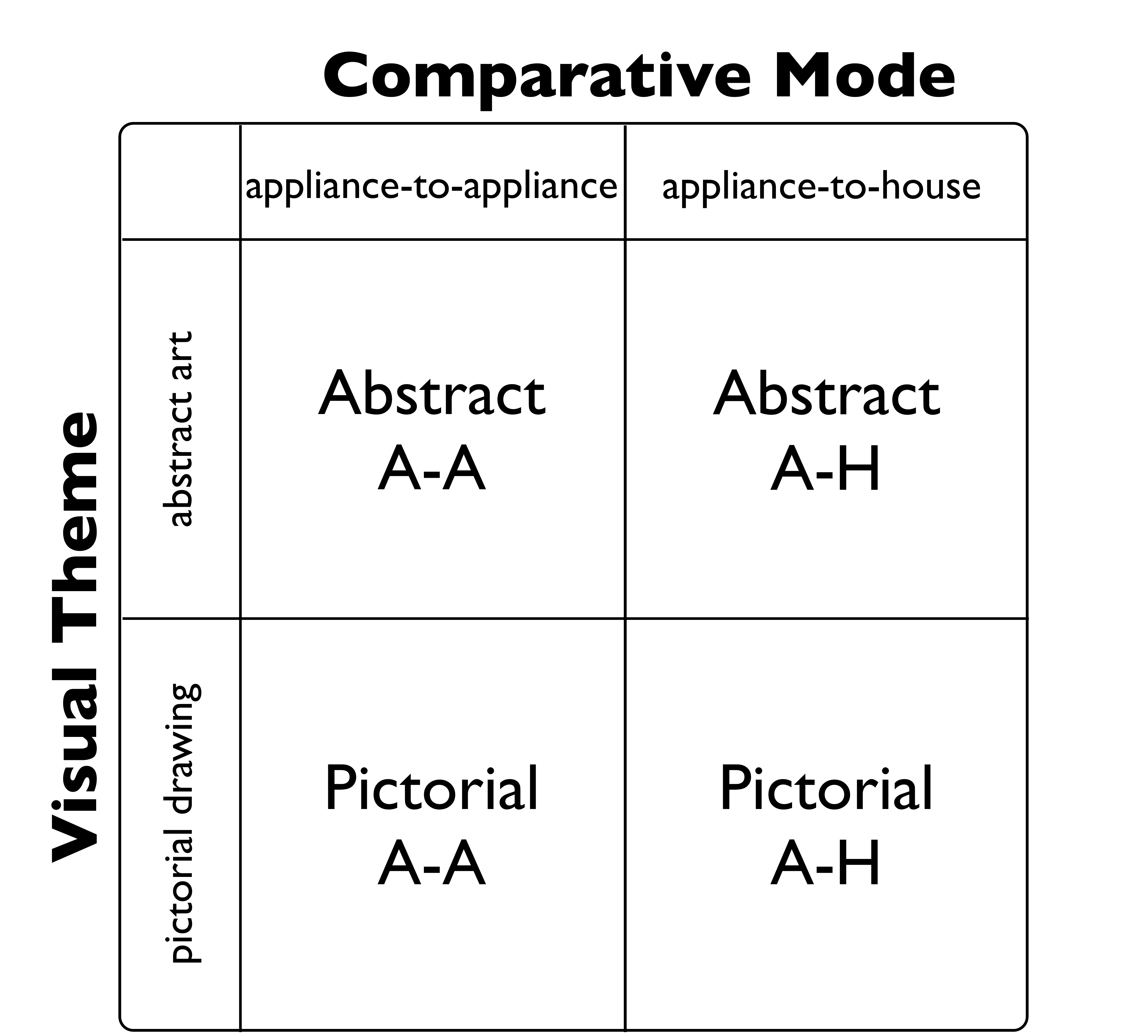}
\caption{Using visual theme and comparative mode to determine the number of eco-visualizations needed.} 
\label{fig:matrix}
\end{figure}

We developed a number of eco-visualizations based on the literature review and our choice of lifestyle factors. Two visual themes were chosen: abstract art and pictorial drawing. Orthogonally, two comparative modes were chosen: comparing appliance-to-appliance energy consumption (A-A) and comparing appliance-to-house energy consumption (A-H; i.e. compare one appliance to the total house energy consumption). This would give us four visualizations to design: abstract art A-A, abstract art A-H, pictorial drawing A-A, and pictorial drawing A-H. This is shown in Figure~\ref{fig:matrix}.

\begin{figure}[ht]
\centering
\includegraphics[width=0.9\columnwidth]{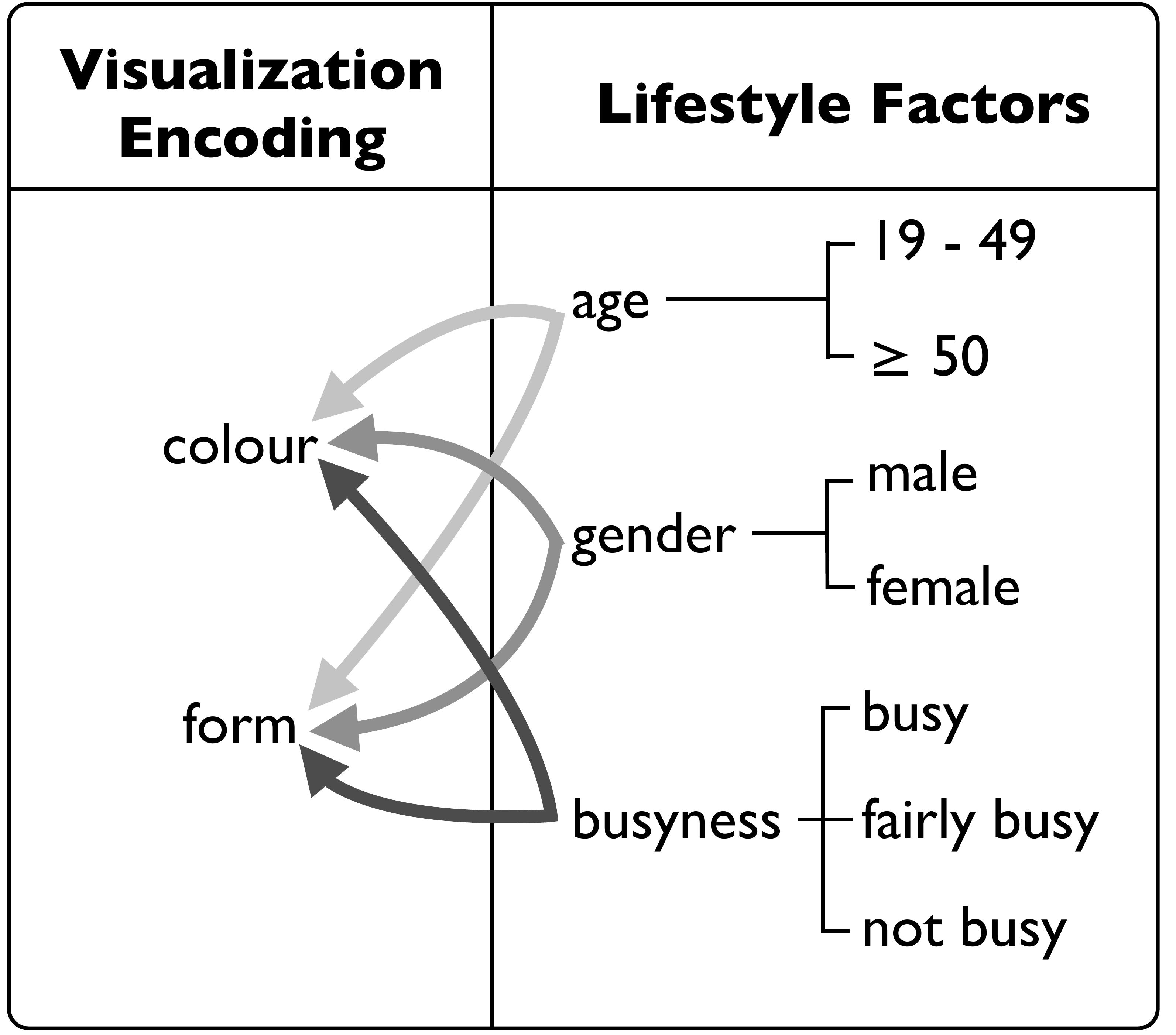}
\caption{How lifestyle factors affect the visualization encoding.} 
\label{fig:enclife}
\end{figure}

Two visualization encoding methods (colour and\break form) were chosen. We wanted to investigate how palette choices (colour) and visualization form (e.g. glyph, radial clock) would be affected by the chosen lifestyle factors (Figure~\ref{fig:enclife}). We then used Pousman's and Stasko's information coding taxonomy (specifically information capacity and representational fidelity) \cite{pousman2006taxonomy} to further focus no how we would design our eco-visualizations.

Information capacity is the representation of the number of ``discrete information sources that a system can represent'' \cite{pousman2006taxonomy}. Some systems can display a single piece of data (ranked \textit{low}) while others display many different information elements on one screen (ranked \textit{high}). Our eco-visualizations rank \textit{medium} because we are visualizing the consumption of a number of appliances over a period of time. While there may be a large number of data points overall, these data points convey the same basic information: an appliance, the relative consumption amount, and a point in time.

Representational fidelity is the display components of a system and ``how the data from the world is encoded into patterns, pictures, words, or sounds'' \cite{pousman2006taxonomy}. Abstract representations (like our eco-visualizations) are considered \textit{low} ranking. 

The two main visualization encoding methods we focused on were colour and form. Colour focused on the selection of a colour palette by providing a couple of palettes for comparison. Form focused on providing different visualization encodings. These visualization encoding methods are discussed in detail in the next two subsections.

\subsubsection{Colour}
There is a large set of literature on the theory behind the use of colour in design. We feel that focusing on a few that emphasize clarity most and support the visual task, yet are aesthetically pleasing, is important. According to Ware, the number of colours that can be used as effective coding is between 6--12 \cite{ware2004information,ware2008visual}. We consider this an important issue when creating visualizations. Other theories that need to be considered are: the use of warm colours to grab attention, the use of similar colours to group closely associated items, and the exclusion of vivid colours that can be perceived as unpleasant and overwhelming \cite{wang2008colour}.

\begin{figure}[ht]
\begin{center}
 \subfigure[Spring Palette]{
 \centering \includegraphics[width=0.9\columnwidth]{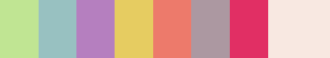}
 \label{fig:spal}
 }
 \subfigure[Autumn Palette]{
 \centering \includegraphics[width=0.9\columnwidth]{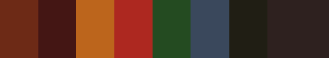}
 \label{fig:fpal}
 }
\end{center}
\caption{Our choice of palettes. Each colour in each palette represents the following (from left to right): house, HVAC, fridge, freezer, oven, TV/PVR, other, and background colour (see Section 4.1).} 
\label{fig:pal}
\end{figure}

For the purpose of our eco-visualizations we have palettes of 7 colours (1 different colour for each appliance monitored). If more appliances were to be monitored then the palette of 7 colours would need to be expanded. If more than 12 appliances were monitored then we would need to consider the idea of grouping similar appliances together either by location (e.g. living room, kitchen) or similarity of function (e.g. kitchen fridge and basement freezer as one group). We considered using warm colours for appliances that potentially have higher consumption to grab attention. We also considered the use of a similar hue for similar appliances (e.g. fridge and freezer). We did not use vivid colours in order to lessen the amount of distraction and perceived unpleasantness. Lastly, we use tools like Vischeck \cite{vischeck} to account for all forms of colour deficiency when we designed our palettes (Figure~\ref{fig:pal}).

Figure~\ref{fig:pal} shows the two palettes we designed: Spring Palette (Figure~\ref{fig:spal}) and Autumn Palette (Figure~\ref{fig:fpal}). Based on opinion that gender may affect colour selection, the Spring Palette was created with slightly lighter more feminine colours and the Autumn Palette was created with slightly darker more masculine colours. User study participants will be able to see each eco-visualization in each palette.

\subsubsection{Form}
Different visual themes should use different visualization forms. For instance, we used \textit{colour supremacy} with \textit{glyphs} for the abstract art visual theme and \textit{24 hour radial clock} with \textit{stacked arcs} for the pictorial drawing visual theme. We are interested in understanding whether or not lifestyle factors affect the choice of the visualization form. At this point we were not sure exactly how choices in visualization form would be affected by lifestyle factors. The user study would need to be designed to take this into account. There may be some visualization forms that are perceived as easier to read than others.

More details about the visualization forms are discussed in our Prototype Design section. An example of colour supremacy is shown in Figure~\ref{fig:leavesaa} and glyphs in Figure~\ref{fig:leavesah}. An example of 24 hour radial clock with stacked arcs is shown in Figure~\ref{fig:spiralaa} and Figure~\ref{fig:spiralah}

\section{Prototype Design}
Our eco-visualizations have been created using Processing, an open source programming language and development environment (processing.org). We developed our own framework (a set of modules) that allowed us to create eco-visualization with ease. This framework handled the gathering of consumption data and other functions that allowed us to focus on developing the visualization code for each eco-visualization. Having this common framework had the added benefit of ensuring that the same data was being visualized for each eco-visualization (using the same web service).  This framework will allow us to easily create more eco-visualizations in the future. The rest of this section is concerned with describing the appliances and the data we used, and describing the eco-visualization that we designed.

\subsection{House and Appliance Data}
Considering the many different appliances that could be found around the house, we chose to visualize data from: HVAC (heat pump), kitchen fridge, basement\break freezer, kitchen oven, and entertainment devices. We have an additional appliance called \textit{other} which is the total house consumption minus the sum of the consumption of all the appliances. 

When the user study was conducted some data came from real meters and other data was carefully mocked up. We have a single house that has two power meters: one to monitor consumption of the whole house, and the other to monitor the consumption of the heat pump (HVAC). Both power meters communicate over Modbus/RS485 and have an existing data collection system  that stores the gathered consumption data in a remote web/database server.

The other appliances have data that is manually mocked up. Meters (using the home automation protocol Insteon) are being installed so we can use real data in later user studies. We gathered the EnerGuide\footnote{Major appliances purchased in Canada come with a sticker that states the expected yearly kWh usage (or energy consumption) of that appliance.} stickers off each appliance and converted their expected yearly consumption into expected hourly consumption. Some appliances we added spikes of consumption. For example, with the oven we added consumption spikes around dinner time. In the case of monitoring entertainment devices (TV/PVR) we looked at an article from BC Hydro\footnote{The provincial crown corporation that is the major electricity supplier for the province of British Columbia, Canada.}~\cite{bchsettopbox} that discussed the the power consumption of the typical DVR/PVR (digital/personal video recorder).

Data was collected once per minute and stored in a remote web/database server. A web service was exposed over the Internet that takes this data and aggregates it into hourly totals. When the web service was called, the last 24 hours of consumption data was returned. The 24 hours of data displayed is a rolling 24 hours and based on current time.

\begin{figure*}[ht]
\begin{center}
 \subfigure[\textbf{Leaves A-A} This visualization renders the consumption from all appliances is rendered in the Autumn Palette. The more one colour is seen the more consumption the appliance associated to that colour has used. In this case the \textit{other} appliance has consumed the most. The other appliance is all the appliances that are not monitored.]{
 \centering \includegraphics[width=\columnwidth]{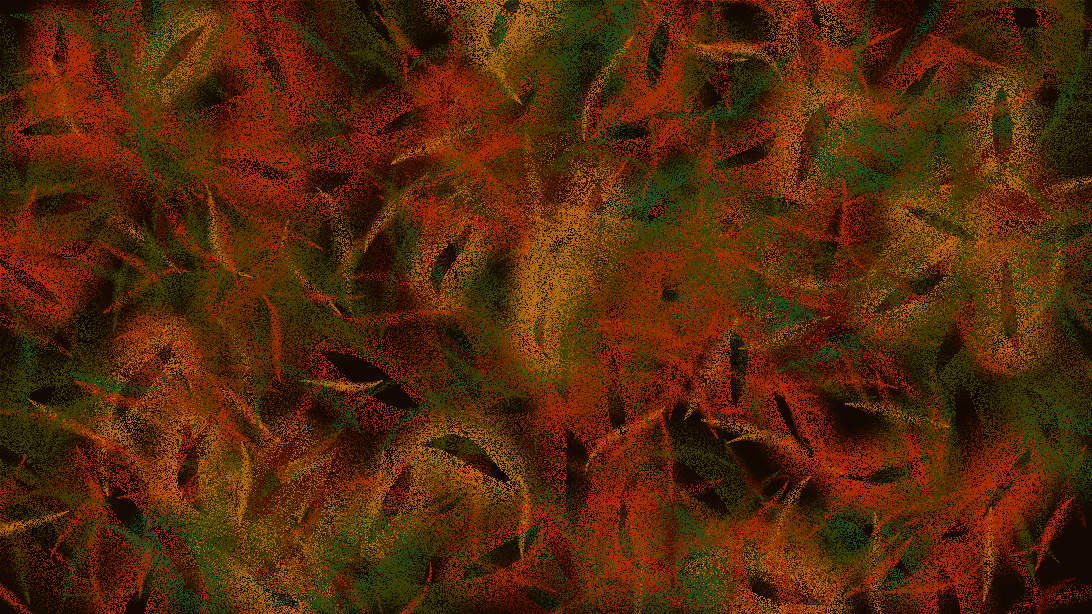}
 \label{fig:leavesaa}
 }
 \subfigure[\textbf{Leaves A-H} With this visualization we are comparing the \textit{other} appliance (the grass at the top of each glyph) and the consumption of the whole house (the ground at the bottom of each glyph). There are 24 glyph (read from left-to-right, top-to-bottom) for 24 hours. The bottom-right is the most current hour.]{
 \centering \includegraphics[width=\columnwidth]{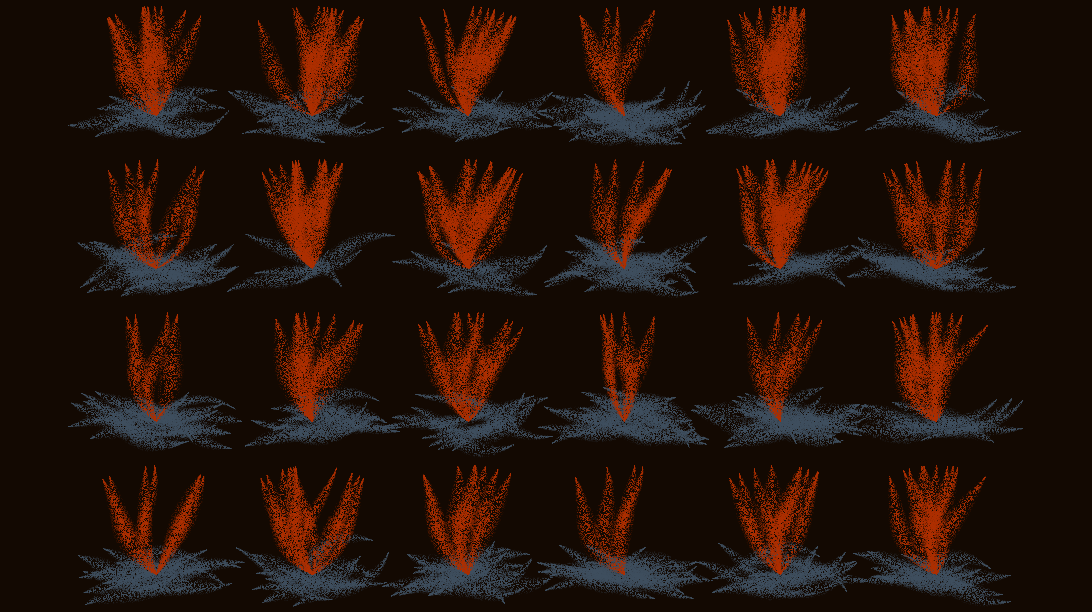}
 \label{fig:leavesah}
 }
 \subfigure[\textbf{Spiral A-A} This visualization compares the consumption from all appliances in the Spring Palette. The order of colours/appliances in based on overall consumption of the last 24 hours, where closer to the center means less consumption. The Spring Palette was used for rendering: green is other appliances, red is HVAC, purple is oven, orange is TV/PVR, yellow is fridge, and blue is freezer.]{
 \centering \includegraphics[width=\columnwidth]{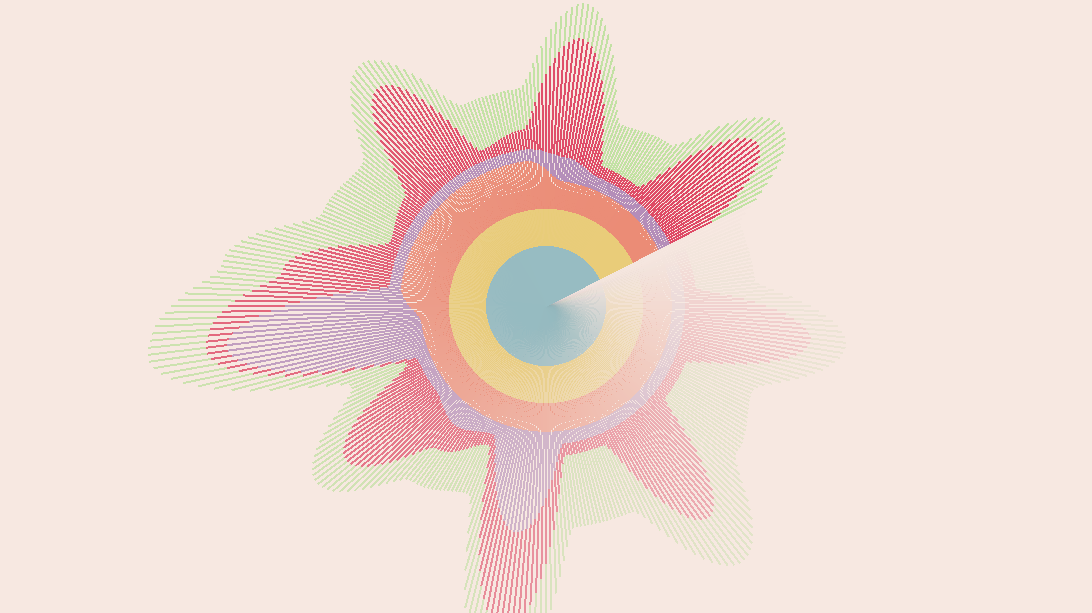}
 \label{fig:spiralaa}
 }
 \subfigure[\textbf{Spiral A-H} With this visualization we are comparing the \textit{other} appliance (center, green) and the consumption of the whole house (outer, gray). The 24 hour radial clock is used with stacked arcs. The length of the arc from center to end is the total consumption for that time period.]{
 \centering \includegraphics[width=\columnwidth]{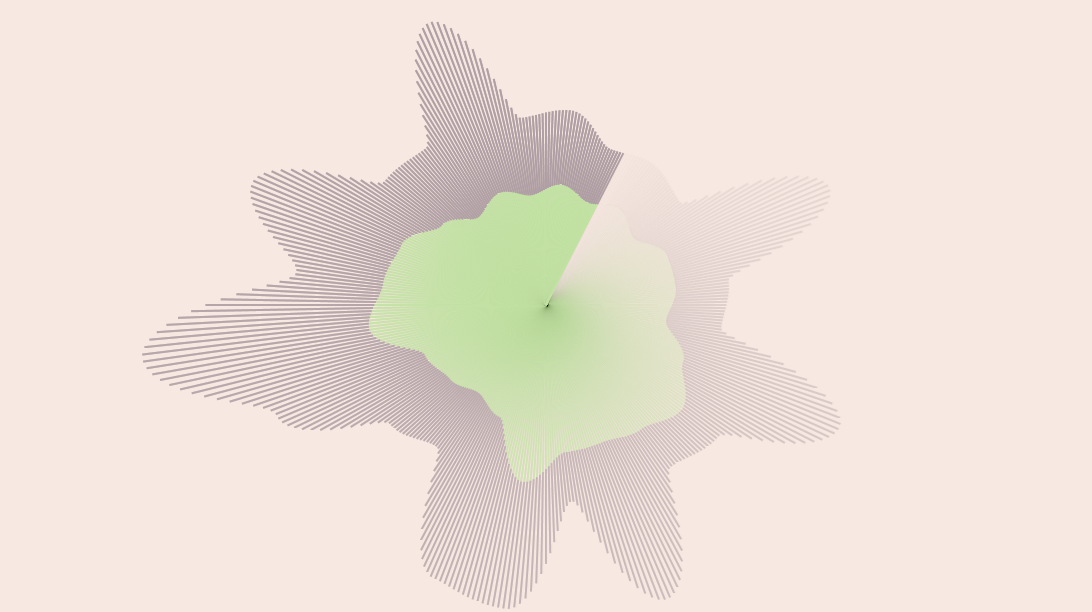}
 \label{fig:spiralah}
 }
\end{center}
\caption{The 4 eco-visualizations based on Figure~\ref{fig:matrix}. Leaves is our abstract art eco-visualization and Spiral is our pictorial drawing eco-visualization.} 
\label{fig:ecovis}
\end{figure*}

\subsection{Leaves of Consumption}
\textit{Leaves of Consumption} (Leaves) has the visual theme of abstract art, and was a modified visualization created by Glassner~\cite{Glassner:2010:PVA:1894991}.
In A-A, Leaves is a visualization that places (at random) coloured leaves over the ambient display for the entire last 24 hours of consumption. Each colour represents a different appliance. Older consumption is drawn first and the screen is refreshed (or redrawn) every 15 minutes. The more colour seen is directly related to the more consumption of the appliance associated to that colour (hence colour supremacy). This visualization is considered abstract art so the consumption data visualized would only give a general sense as to the consumption amount of each appliance. A screen capture of this can be seen in Figure~\ref{fig:leavesaa}. 

In A-H, Leaves uses glyphs. One glyph is drawn for each hour (a total of 24) starting at the top left (hour 24) and ending at the bottom right (now). Each row being read from left to right. The glyphs resemble tufts of grass. The ground (bottom) is the house consumption and the grass (top) is the consumption of the selected appliance. Each colour represents a different appliance. If the glyph has more ground and less grass that would mean the appliance consumed less energy. If the glyth has more grass and less ground that would mean the appliance consumed more energy. A screen capture of this can be seen in Figure~\ref{fig:leavesah}. 

\subsection{Spiral of Consumption}
\textit{Spiral of Consumption} (Spiral) has the visual theme of pictorial drawing, and was a visualization inspired by Obieta's designs \cite{obieta2011dia04}. Spiral uses a 24 hour radial clock with stacked arcs in both A-A and A-H comparative modes. 
The 24 hour radial clock is slowly refreshed (about 4 times/hour) by its hand moving in a clockwise manner. Hour 24 would be in the top vertical position (12 on a regular clock) and hour 12 would be on in the bottom vertical position (6 on a regular clock). 

In A-A, appliances are sorted in the order of amount of consumption. Each arc of the radial clock is the length of the sum of consumption of all the appliances. Each arc has colours of different lengths. Each colour represents a different appliance and the length in the amount of consumption for that appliance. A screen capture of this can be seen in Figure~\ref{fig:spiralaa}. A-H works the same way except only one appliance and the house is shown. Meaning, each arc is made up of two colours, one for the appliance and the other for the house. A screen capture of this can be seen in Figure~\ref{fig:spiralah}.

\section{User Study}
Participants sat in front of a 15.5-inch anti-glare MacBook Pro with a screen resolution of $1680 \times 1050$ pixels. We started each session with a brief introduction of our study and what they were about to see. A demonstration of the eco-visualization lasted for 15 minutes. This was followed by the completion of a questionnaire and an interview. The total session took about 20 minutes to complete.

For each session we sat with the participant and demonstrated each of the 4 eco-visualizations, one at a time. We first presented Spiral (in A-A and then in A-H) followed by presenting Leaves (in A-A and then in A-H). This was demonstrated the same way for all participants. The demonstration of each eco-visualization also included an explanation of how to interpret the screens. Participants could ask questions about the eco-visualizations at any time during the demonstration. 

\subsection{Method}
To explore the control and balance of lifestyle and visualization factors we used scientific method to create an informal user study. At this point we still need to explore the ecological validity of these factors. It is important to note that a formal user study would only be appropriate is these factors have already been identified \cite{tory2005doeswork,Greenberg:2008:UEC:1357054.1357074,acevedo2006subjective}.

\subsection{Hypotheses}
We further narrowed down our conjecture to a number of hypotheses about our user study. They are listed below:
\begin{description}
\item[\textbf{H1}] Females will like more feminine colours (Spring Palette) and males will like more masculine colours (Autumn Palette).
\item[\textbf{H2}] Those who perceive themselves as busy will want an eco-visualization designed to be easier to read.
\item[\textbf{H3}] A majority of participants will pick the living room as the location to place our eco-visualizations.
\end{description}

\subsection{Coding and Measurement}
The questionnaire and interview contained both coded and open-ended questions. Coded questions where measured as follows:
\begin{description}
\item[\textbf{Age}] Choose the participant's age group ($19-49$ or $\geq 50$).
\item[\textbf{Gender}] Choose the participant's gender (male or female).
\item[\textbf{Busyness}] Choose the participant's perceived level of busyness (busy, fairly busy, or not busy). Where Busy was on one side of the scale and Not Busy was on the other side. Fairly Busy was in the middle.
\item[\textbf{Visual Theme Preference}] Choose which theme was preferred (Leaves or Spiral).
\item[\textbf{Comparative Mode Preference}] Choose which mode was preferred(A-A or A-H).
\item[\textbf{Colour Palette Preference}] Choose which palette was preferred (Spring or Autumn).
\item[\textbf{Preferred Location}] Choose any number of 9 locations (Kitchen, Bathroom, Bedroom, Living/Family Room, Dining Room, Office, Entrance, Hallway, or Other). This is the location that participants would prefer to have an ambient display.
\end{description}

\subsection{Setting and Location}
User study sessions ran at two locations: a home, and a university lab.
The home location had 18 participants and the university lab had 6 participants. In both locations, the source of artificial light in the room was casting from behind the computer screen. 

\subsection{Participants}
Our user study included 24 participants recruited from a network of friends, family, and university students. The sampling method was not random, 6 males and 6 females within the age range of $19-49$ and additional 6 males and 6 females within the age range of $\geq 50$ were purposely chosen. This was done to match our choice of independent variables (age and gender).

\begin{table}
\begin{center}
\begin{tabular}{lcccc}
  && \multicolumn{3}{c}{Comparative Mode} \\
  \cline{3-5}\noalign{\smallskip}
  Busyness && A-A & A-H & Both \\
  \hline\noalign{\smallskip}
  Busy        &&  2 & 10 &  5 \\
  Fairly Busy &&  2 &  3 &  1 \\
  Not Busy    &&  0 &  1 &  0 \\
  \hline\noalign{\smallskip}
  Totals      &&  4 & 14 &  6 \\
  \hline\noalign{\smallskip}
\end{tabular}
\end{center}
\caption{Number of participants broken down by comparative mode preference versus level of busyness. Of the participants studied, the majority considered themselves busy and preferred the A-H comparative mode.} 
\label{table:cmbusy} 
\end{table}

\section{Results Summary}
The data for each session was collected electronically, stored, and correlated on a secure server. Data collected during the study was analyzed using different methods. Multiple answer questions were analyzed based on the frequency of answers given. With regards to the descriptive questions, a generic approach to qualitative analysis was taken, with open coding leading to the development of themes in the observation and interview data.

When looking at participants who perceived themselves as \textit{busy} (Table~\ref{table:agegender}), 70\% (17 or 24) classified themselves as such. The majority of these participants were either males between the ages of $19-49$ or females $\geq 50$ year of age.

The A-H comparative mode was preferred by $2/3$ of the participants (Table~\ref{table:agegender}), where 14 participants selected A-H only and 6 participants selected both (A-A and A-H). Results are the same for participants who perceived themselves as \textit{busy} (Table~\ref{table:cmbusy}).

When it came to colour palette preferences (Table~\ref{table:agegender}), 50\% of participants preferred Spring, the other 50\% preferred Autumn. Ages $19-49$ generally had a tendency towards Spring colours while $\geq 50$ leaned more towards Autumn colors (Table~\ref{table:agegender}).

For visual theme (Table~\ref{table:agegender}) Spiral was preferred by 83\% (20 of 24). All male participants preferred Spiral rather than Leaves.

Finally, location preference (Table~\ref{table:agegender}) was examined. Out of the 24 participants 64 selections were made. No participants chose all or none of the locations. Kitchen (16 selections) and Living/Family Room (15 selections) lead with a wide margin. Next highest preferred location was Office at 7 selections. Participants who perceived themselves as \textit{busy} also preferred Kitchen and Living/Family Room locations.

\begin{table*}
\begin{center}
\begin{tabular}{llcccccccccccccc}
  & &~~~& \multicolumn{3}{c}{Comparative Mode} &~~~& \multicolumn{2}{c}{Colour Palette} &~~~& \multicolumn{2}{c}{Visual Theme} &~~~& \multicolumn{3}{c}{Busyness} \\
  \cline{4-6}\cline{8-9}\cline{11-12}\cline{14-16}\noalign{\smallskip}
  Age & Gender && A-A & A-H & Both && Autumn & Spring && Leaves & Spiral && Busy & Fairly & Not \\
  \hline\noalign{\smallskip}
  $19-49$      & Female &&  1 &  5 &  0 &&  4 &  2 &&  3 &  3 &&  3 &  3 & 0 \\
               & Male   &&  0 &  5 &  1 &&  1 &  5 &&  0 &  6 &&  6 &  0 & 0 \\
  \hline\noalign{\smallskip}
  $\geq 50$    & Female &&  1 &  2 &  3 &&  5 &  1 &&  1 &  5 &&  5 &  1 & 0 \\
               & Male   &&  2 &  2 &  2 &&  2 &  4 &&  0 &  6 &&  3 &  2 & 1 \\
  \hline\noalign{\smallskip}
  Totals       &        &&  4 & 14 &  6 && 12 & 12 &&  4 & 20 && 17 &  6 & 1 \\
  \hline\noalign{\smallskip}
\end{tabular}
\end{center}
\caption{Number of participants broken down by comparative mode preference, color palette preference, visual theme preference, and busyness versus age and gender.
Of the participants studied, the majority preferred the A-H comparative mode, and were evenly split on colour palette preference. The majority of participants preferred the Spiral visual theme, and a majority considered themselves busy.} 
\label{table:agegender} 
\end{table*}

\begin{table*}
\begin{center}
\begin{tabular}{llccccccccc}
  & & \multicolumn{9}{c}{Preferred Location} \\
  \cline{3-11}\noalign{\smallskip}
  Age & Gender & Kitchen & Living/Family & Office & Bedroom & Entrance & Hallway & Bathroom & Dining & Other \\
  \hline\noalign{\smallskip}
  $19-49$      & Female & 5 & 5 & 1 & 2 & 0 & 1 & 2 & 1 & 0 \\
               & Male   & 3 & 5 & 3 & 4 & 4 & 2 & 1 & 2 & 1 \\
  \hline\noalign{\smallskip}
  $\geq 50$    & Female & 6 & 3 & 1 & 0 & 1 & 2 & 0 & 0 & 0 \\
               & Male   & 2 & 2 & 2 & 0 & 1 & 0 & 1 & 0 & 1 \\
  \hline\hline\noalign{\smallskip}
  \multicolumn{11}{l}{Busyness} \\
  \hline\noalign{\smallskip}
  \multicolumn{2}{l}{Busy}        & 12 & 10 & 5 & 4 & 6 & 5 & 2 & 2 & 2 \\
  \multicolumn{2}{l}{Fairly Busy} &  4 &  5 & 1 & 2 & 0 & 0 & 2 & 1 & 0 \\
  \multicolumn{2}{l}{Not Busy}    &  0 &  0 & 1 & 0 & 0 & 0 & 0 & 0 & 0 \\
  \hline\hline\noalign{\smallskip}
  \multicolumn{2}{l}{All Participants} & 16 & 15 & 7 & 6 & 6 & 5 & 4 & 3 & 2 \\
  \hline\noalign{\smallskip}
\end{tabular}
\end{center}
\caption{Number of participants broken down by preferred location (for display within the house) versus: first, by age and gender; then by busyness; and finally, by all participants. Kitchen and Living/Family Room is the preferred choice of the participants studied.} 
\label{table:prefloc} 
\end{table*}

\section{Discussion}
The results of our study suggest there is a trend that shows an overwhelming number of \textit{busy} participants preferred the A-H comparative mode (Table~\ref{table:cmbusy}). This might suggest that the A-A comparative mode has too much information to glean from an eco-visualization. With A-H, participants could \textit{better judge the relative amounts from a distance}, and \textit{they needed less time to dwell on it}. The fact that the participants also preferred Spiral over Leaves (Table~\ref{table:agegender}) further suggests that eco-visualizations need to be simple and easy to understand. As one participant put it: \textit{the leaves is more beautiful, like a painting but I can understand the information in Spiral better}. And from another: \textit{I think the spiral look works better, this is about information, the art'y leaves don't display information very good}. All this shows support for our hypothesis \textbf{H2} and (we believe) begins to answer our point on \textit{effective feedback}.

Our study suggests that age, gender and busyness seemed irrelevant to the visual theme (Table~\ref{table:agegender}). However, demographics indicated that people with an Art background were the participants who selected the\break Leaves display. As one participant with an art background commented: \textit{Living room 'cause its like a painting you can look at it occasionally}. We now think that occupation affects the choice of visual theme, but we need to investigate this further.

With the general assumption that men prefer sharp\-er, darker colours and women lean more towards softer colours we demonstrated our two Autumn and Spring palettes. There was no distinguishing difference between male and female, but there was a difference in age group. Ages $19-49$ generally had a tendency towards Spring colours while $\geq 50$ leaned more towards Autumn colors (Table~\ref{table:agegender}). This clearly shows there is no support for hypothesis \textbf{H1}. We now believe that colour palette preference is more linked to a participant's age not gender. A further literature review on this matter has found studies that support this observation \cite{dittmar2001changing,leecolor}.

Preferred location in a house for ambient displays came down to two main locations: Kitchen and Living/Family Room. This suggests that there is support for our hypothesis \textbf{H3}. We would also note that these results confirm the findings that were presented in a recent paper by Rogers and Bartram \cite{6065016}.

\subsection{Design Implications}
Further postulation on the broader meaning of our user study results have motivated us to comment further on our findings. Before we elaborate we would cite a relevant comment that supports what we are about to discuss.

\begin{quote}
``Displays are regarded as solitary objects - only the relationship between observer and display is taken into account. Yet, the relationship between a display and its context is equally important for the experience, especially when the display is seamlessly embedded into the public, architectural environment'' \cite{moere2009beyond}.
\end{quote}

We have found a need to create an actual ambient display that is stand-alone, that can be hung on a wall, and is not a computer nor computer monitor. For example, using a wood frame around the LCD display. We have also found that any future user study should be conducted within the participant's home. By doing so, we begin to address the context problem Moere \cite{moere2009beyond} identifies.
Participants are most comfortable in their homes and with the ability of placing the ambient display in different locations around the home, users do not need to imagine the display in different places. They actually get to see it in the different locations around the home. 
We need to consider that ambient displays should be of different forms and different sizes; and this might depend of the location within the home of where the ambient display is. And yes, they can be screen savers, too.

It is our intent to provide users with information on how they consume energy through tacit means. Homeowners who feel that their energy consumption is too high (say for financial reason) may wish to modify how (the amount or when) they consume. This modification is directly tied to the lifestyle factors that we have identified. Strategies on how to consume less, based on a homeowner's lifestyle factors, is still an open research question that we are studying.

\subsection{New Hypothesis}
All this has led us to a new hypothesis. We anecdotally noted that participants who were artists preferred Leaves. \textit{We now think that occupation affects the choice of visual theme.}

\section{Conclusions and Future Work}
We have provided a number of eco-visualizations that were then subjected to a user study. From our user study we have observed a number of things. People who are busy want displays that can be understood at a glance. We found that the context of the ambient display (eco-visualization) matters. This means that in-home user studies need to be looked at. More importantly, we found that most participants liked our eco-visualizations and wanted a ``product'' like this for their homes. 

Our future work includes developing an in-home study and conducting it with a large number of homeowners. We are also interested in what younger participants (ages $6-19$) would say. Also, we would be providing more choices in terms of visual themes and provide additional colour palettes to choose from.


\section*{Acknowledgments}
This work was supported in part by the Graphics, Animation and New Media (GRAND) Network of Centres of Excellence of Canada.


\bibliographystyle{spmpsci}
\bibliography{z-ref}

\end{document}